# Photonic Tensor Cores for Machine Learning

Mario Miscuglio[1], Volker J. Sorger[1,*]

[1]Department of Electrical and Computer Engineering, George Washington University, Washington, DC 20052, USA
Email: *sorger@gwu.edu

**Abstract:**

*With an ongoing trend in computing hardware towards increased heterogeneity, domain-specific coprocessors are emerging as alternatives to centralized paradigms. The tensor core unit has shown to outperform graphic process units by almost 3-orders of magnitude enabled by higher signal throughout and energy efficiency. In this context, photons bear several synergistic physical properties while phase- change materials allow for local nonvolatile mnemonic functionality in these emerging distributed non van-Neumann architectures. While several photonic neural network designs have been explored, a photonic tensor core to perform matrix vector multiplication and summation is yet to be implemented. In this manuscript, we introduce an integrated photonics-based tensor core unit by strategically utilizing i) a photonic parallelism via wavelength division multiplexing, ii) high 2 Peta-operations-per second throughputs enabled by 10's of picosecond-short delays from optoelectronics and compact photonic integrated circuitry, and iii) near-zero power-consuming novel photonic multi-state memories based on phase-change materials featuring vanishing losses in the amorphous state. Combining these physical synergies of material, function, and system, we show, supported by numerical simulations, that the performance of this 4-bit photonic tensor core unit can be one order of magnitude higher for electrical data, whilst the full potential of this photonic tensor processor is delivered for optical data being processed, where we find a 2-3 orders higher performance (operations per joule) as compared to an electrical tensor core unit whilst featuring similar chip areas. This work shows that photonic specialized processors have the potential to augment electronic systems and may perform exceptionally well in network-edge devices in the looming 5G networks and beyond.*

## I. Introduction

Aiming to replicate brain functionalities remains a captivating challenge, which does not only aspire and intrigue human feats, but also has shown to provide technological usefulness for modern societies. Indeed, Machine Learning (ML), performance by Neural Networks (NN), has become a popular approach to Artificial Intelligence (AI), and consists of training a system to learn how to perform unsupervised decision classifications on unseen data; once a NN is trained, it can be implemented to produce an *inference*, in other words recognizing and classifying objects or patterns.

Most NNs unravel multiple layers of interconnected neurons/nodes. Each neuron and layer, as well as the network interconnectivity, is essential to perform the task which the network has been trained for. In their connected layer, NNs strongly rely on vector matrix math operations[1], in which large matrices of input data and weights are multiplied, according to the training. Complex multi-layered deep NNs, in fact, require a sizeable amount of bandwidth and low latency for satisfying the vast operation required for performing large matrix multiplication (MM) without sacrificing efficiency and speed[2].

Since the dawn of the computing era, due to the ubiquity of matrix math, which extends to neuromorphic computing, researchers have been investigating optimized ways to efficiently multiply matrices[3–6]. To corroborate this statement, engineering a platform which performs energy efficient and faster matrix multiplication enables solving linear algebraic problems, such as inverting matrices, systems of linear equations, and finding determinants. Even some basic graph algorithms[7] are obstructed by the speed at which matrix multiplication is computed.

For a general-purpose processor offering high computational flexibility, these matrix operations take place serially, one-at-a-time, while requiring continuous access to the cache memory, thus generating the so called "von Neumann bottleneck". Specialized architectures for NNs such as Graphic Process Units (GPUs) and Tensor Process Units (TPUs), have been engineered to reduce the effect of the von Neumann bottleneck enabling cutting-edge machine learning models. The paradigm of these architectures is to offer domain-specificity such as being optimized for convolutions or Matrix Vector Multiplications performing operations, unlike CPUs, in parallel deploying a systolic algorithm.

GPUs have thousands of processing cores optimized for matrix math operations, providing tens to hundreds of TFLOPS (Tera FLoating Point OPerations) of performance which makes GPUs the obvious computing platform for deep NN-based AI and ML applications. GPUs and TPUs are particularly beneficial with respect to CPUs, but when used to implement deep NN performing inference on large 2-dimensional data sets such as images, they are rather power-hungry and require longer computation time (> tens of ms). Moreover, smaller matrix multiplication for less complex inference tasks (e.g. classification of handwritten digits of the Modified National Institute of Standards and Technology database - MNIST[8]) are still challenged by a non-negligible latency predominantly due to the access overhead of the various memory hierarchies and the latency in executing each instruction in the GPU[9].

Given this context of computational hardware for obtaining architectures that mimic efficiently the biological circuitry of the brain, it is necessary to explore and reinvent the operational paradigms of current logic computing platforms when performing matrix algebra, by replacing sequential and temporized operations, and their associated continuous access to memory, with massively parallelized distributed analog dynamical units, towards delivering efficient post-CMOS devices and systems summarized as non von Neumann architectures. In this paradigm shift the wave nature of light and related inherent operations, such as interference and diffraction, can play a major role in enhancing computational throughput and concurrently reducing the power consumption of neuromorphic platforms. In recent years, the revolutionizing impact of NNs contributed to the development of a plethora of emerging technologies, ranging from free space diffractive optics[10] to nanophotonic processors[11–15] aiming to improve the computational efficiency of specific tasks performed by NN. Integrated photonic platforms can indeed provide parallel, power-efficient and low-latency computing, which is possible because analog wave chips can a) perform the dot product inherently using light matter interactions such as via a phase shifter or modulator, b) enable signal accumulation (summation) by either electromagnetic coherent interference or incoherent accumulation through detectors, and c) enable parallelism strategies and higher throughput using multiplexing schemes.

Additionally, we firmly believe, assisted by state-of-the-art theoretical framework[16], that future technologies should perform computing tasks in the domain in which their time varying input signals lay, exploiting their intrinsic physical operations. In this view, photons are an ideal match for computing node-distributed networks and engines performing intelligent tasks over large data at the edge of a network (e.g. 5G), where the data signals may exist already in the form of photons (e.g. surveillance camera, optical sensor, etc …), thus pre-filtering and intelligently regulating the amount of data traffic that is allowed to proceed downstream towards data centers and cloud systems[17].

Here, we explore a photonic tensor core (PTC) able of performing 4x4 matrix multiplication and accumulation with a trained kernel in one-shot (i.e. non-iterative) and entirely passive; that is, once a NN is trained, the weights are stored in a 4bit multilevel photonic memory directly implemented on-chip, without the need of neither additional electro-optic circuitry nor off-chip DRAM. The photonic memories feature low-losses phase-change nanophotonic circuits based on wires of $G_2Sb_2Se_5$ deposited on a planarized waveguides which can be updated using electrothermal switching and read all-optically. Electrothermal switching is enabled by tungsten heating electrodes which clamps the Phase Change Memory (PCM) wire.

This work represents the first approach towards the realization of a photonic tensor processor which could scale the number of multiply-accumulate (MAC) operations by several orders of magnitude while significantly suppressing power consumption and latency compared to the state-of-the-art hardware accelerators.

## II. RESULTS AND DISCCUSION
## II.1 Matrix multiplication algorithms

Considering a naïve ('schoolbook') algorithm, a multiplication between two square matrices, each with $n \times n$ entries, is characterized by a computational complexity of $O(n^3)$ (**Fig, 1**). Which means that the total operations required for performing this operation scales cubically with the size ($n$) of the square matrix. Even for optimized algorithms, such as Strassen[18] or Winograd[19], the complexity of the algorithm still requires a *$O(n^{2.373})$ (Fig. 1a)*. Such 'flat' computational complexity scaling requires long latency when computed in regular CPUs since operations are executed sequentially at each clock cycle. A tensor core unit algorithm is an interesting alternative, however not for a reduced computational (operation) complexity, which is indeed still $O(n^3)$, but because of its capability of exploiting parallel architectures and a systolic algorithm[20]; for instance, it can implement a multiplication between two $n \times n$ matrices with an operational time complexity of $O(n^2)$, primarily dominated by reading/writing the input and output matrices, even if the number of total operations is still $O(n^3)$.[21] In other words, one must distinguish between the complexities of the computational algorithm versus that of the system's execution time. In this regard, the computational complexity scaling results indeed suggest that the main focus should be placed on the fundamental improvements in the time complexity, thus diverting the focus onto hardware such as architectural, circuit, and component-level novelties including new device physics to speed-up the matrix processor's signal 'rate', or in exploiting parallelization strategies (**Fig. 1**). Interestingly, both suggest to consider optics or integrated photonics given a) the various physical multiplexing options (e.g. spectral, mode, polarization, etc...), and b) the short delay in the range of 1-10's of picoseconds from modern optoelectronics devices with 3-dB bandwidths of 10's GHz in photonic foundries [Interuniversity Microelectronics Centre - IMEC, American Institute for Manufacturing Integrated Photonics - AIM] and even approaching 100's of GHz in laboratory settings.[22]

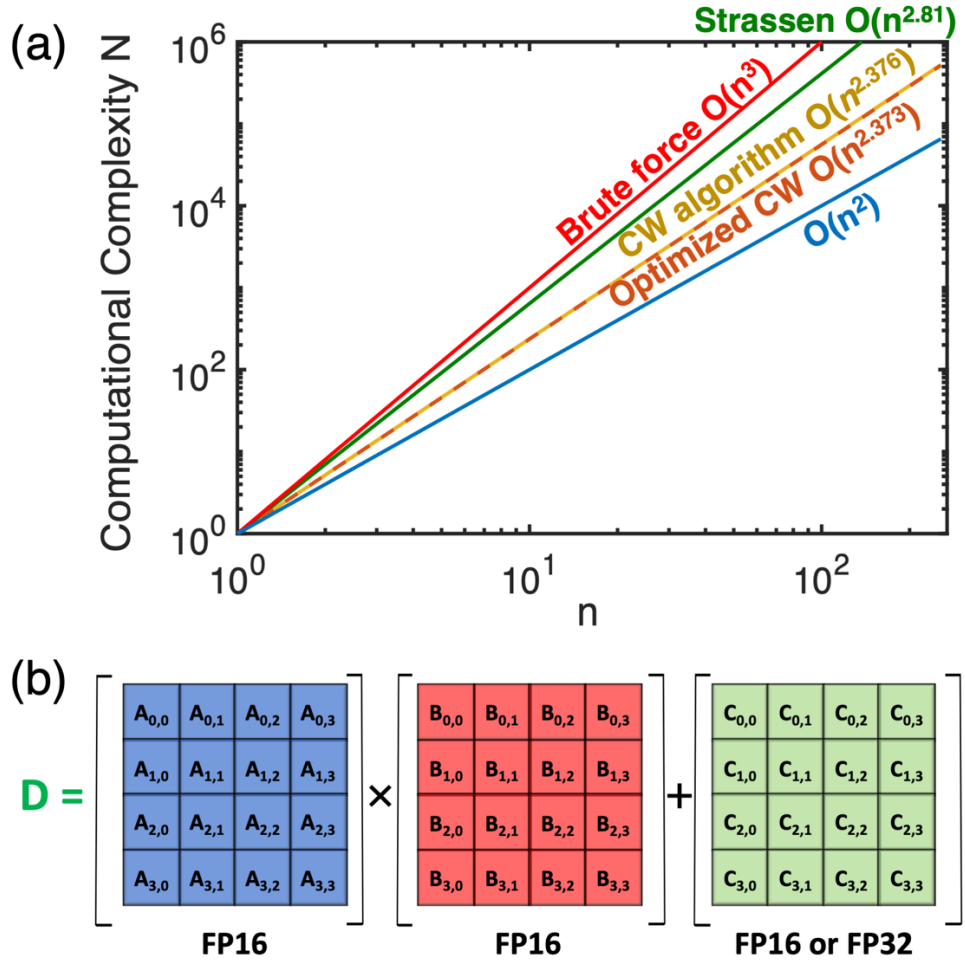

*Figure 1. **Matrix Multiplication.** (a) Computational complexity of a n × n matrices multiplication for different algorithms: Brute-force (naïve), Strassen, Coppersmith–Winograd, Optimized CW and the longed $O(n^2)$ algorithms when performed on GPUs TCUs. A multiplication between two square matrices, each with n × n entries, is characterized by a computational complexity of $O(n^3)$. This means that the total operations required for performing this operation scales cubically with the size (n) of the square matrix. For optimized algorithms, such as Strassen, or Winograd, the complexity of the algorithm still requires a $O(n^{2.373})$. This 'flat' computational complexity scaling requires long latency when computed in regular CPUs since operations are executed sequentially at each clock cycle. A tensor core unit algorithm is an interesting alternative, however not for a reduced computational (operation) complexity, which is indeed still $O(n^3)$, but because of its capability of exploiting parallel architectures and a systolic algorithm; (b) 4x4 Matrix multiplication and accumulation performed by a GPU tensor core. Tensor cores are used to perform large-scale 2-dimensional, or higher dimensional, matrix operations built up from smaller elements, namely Tensor Core Units (TCUs). Each TCU operates on a 4x4 matrices and performs the following operation: D = A×B + C where A, B, C, and D are 4x4 matrices. The matrix multiply inputs A and B are FP16 matrices, while the accumulation matrices C and D may be FP16 or FP32 matrices.*

For this reason, tensor cores are used to perform large-scale 2-dimensional, or higher dimensional, matrix operations built up from smaller elements, namely Tensor Core Units (TCUs). Each TCU operates on a 4x4 matrices and performs the following operation: ***D = A×B + C*** where ***A***, ***B***, ***C***, and ***D*** are 4x4 matrices. The matrix multiply inputs A and B are FP16 matrices, while the accumulation matrices C and D may be FP16 or FP32 matrices (**Fig. 1b**).[23]

In order to significantly reduce the execution time of large matrix multiplication or to avoid the non-negligible latency given by the time to collect data from a specific TCU, here we present a design based on silicon photonic TCUs performing matrix multiplication inherently with its latency time-of-flight only limited by the delay of the detection mechanism, which is well below 10's of ps short in modern high-speed photoreceivers[24–26].

Unlike digital electronics which rely on logic gates, in integrated photonics, multiply and accumulation, and more in general linear algebraic operations can be performed inherently and non-iteratively benefitting from the intrinsic parallelism provided by the electromagnetic nature of the signals and efficient light matter interaction. In this regards, integrated photonics is an ideal platform for mapping specific complex operations one-to-one into hardware, and in some cases algorithms, achieving time complexity (**O**(1)). This allows processing information that does not scale with the number of elementary operations nor input size with latency given only by the time-of-flight of the photon in the photonic chip and detection mechanism, assuming sufficiently available parallel hardware/wavelengths. Note that this is valid when a rapidly varying input is multiplied by a fixed or seldomly changing matrix (kernel), which is the case in machine learning when performing inference tasks. In the next section (II.2), we describe an architecture, namely PTC Unit that is homomorphically mapping an (exemplary) 4x4 matrix multiplication with a fixed (stored) kernel into a photonic platform, with a time complexity of **O**(1) and running time of < 20 ps (considering Photonic Foundry available 50 GHz Photodetector[25]).

## II.2 Photonic Tensor Core architecture

The main advantage of performing a matrix multiplication and accumulation operations in the photonic domain is that they can be performed with near-zero static power consumption (i.e. static or rarely changing kernel), while allowing for low latency, given only by the time of flight of photon.

To build a PTC, we use 16 fundamental units, namely dot-product engines, which perform an element-wise multiplication whilst featuring a Wavelength Division Multiplexing (WDM) scheme, similarly to [28–30], for parallelizing the operation (**Fig. 2a**).

The dot product engine (**Fig. 2b**) performs the multiplication between two vectors, namely, between the $i^{th}$ row of the input matrix ***A*** and the $j^{th}$ column of the kernel ***B***. In this scheme, the $i^{th}$ row of the input matrix is given by WDM signals which, if not already in the optical domain, are modulated by high-speed (e.g. Mach Zehnder) modulators. The $j^{th}$ column of the kernel matrix is loaded in the photonic memory by properly setting its weight states. Availing light-matter interaction with the phase-change memory, the inputs, opportunely spectrally filtered by microring resonators (MRR), are weighted in a seemingly quantized electro-absorption scheme (i.e. amplitude modulation), thus performing element-wise multiplication. The element-wise multiplications are thence incoherently summed up using a photodetector, which amounts to a MAC operation (***D**_{ij}*).

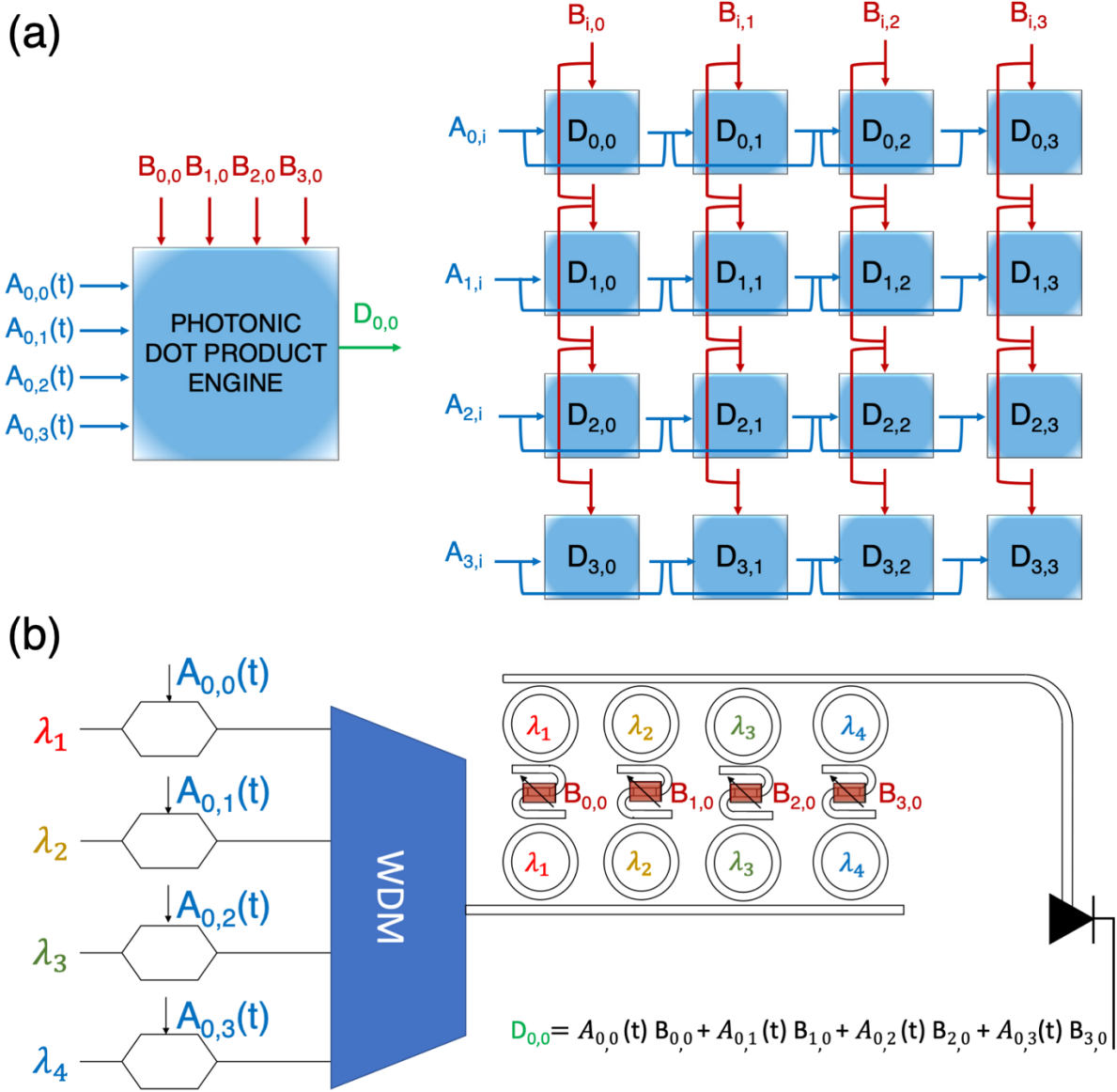

*Figure 2. Photonic Tensor Core (PTC) and Dot-product Engine. (a) The PTC is constituted by 16 dot product engine which inherently and independently performs row by column pointwise multiplication and accumulation. (b) The dot product engine performs the multiplication between two vectors, namely, between the i-th row of the input matrix A and the j-th column of the kernel B. Here, the i-th row of the input matrix is given by WDM signals which are modulated by high-speed (e.g. Mach Zehnder) modulators. The j-th column of the kernel matrix is loaded in the photonic memory by properly setting its weight states. Availing light-matter interaction with the phase- change memory, the inputs, opportunely spectrally filtered by microring resonators (MRR), are weighted in a seemingly quantized electro-absorption scheme (i.e. amplitude modulation), thus performing element-wise multiplication. The element-wise multiplications are thence incoherently summed up using a photodetector, which amounts to a MAC operation ($D_{ij}$).*

It is worth noticing that contrary to other photonic NN implementations[13,15,29,31] based on micro-ring modulators, the transmission of the microrings is not actively tuned for performing filtering, but just utilized for passively selecting frequency to be modulated by the photonic memories. This allows to have more control on the inter-channel crosstalk and potentially extending the number of wavelengths in a Dense WDM (DWDM) scheme without being affected by the induced quality factor variation caused by the variation of absorption coefficient. Additionally, for each couple of micro-ring resonators, our architecture comprises low loss programmable multi-state photonic memories (discussed in section II.3), which unlike electro-optic modulators can retain information without any static power consumption and do not add considerable losses.

### II.3 Photonic Memories

The benefits given by the intrinsic electromagnetic nature of the signals can potentially be hindered by the optoelectrical and electrooptical transductions, as well as by the repeated access to a digital and nonvolatile memory, which impacts on the overall operation speed, while producing considerable additional energy loss. For this reason, having a heterogeneously integrated optimized photonic memory which retains information in a non-volatile fashion, poses a great advantage, especially when implementing NN performing inference, where the trained weights are only rarely update (i.e. depending on the application daily, monthly, yearly, if ever). To provide this functionality, a multi-state photonic memory device, which comprises of multiple $Ge_2Sb_2Se_5$ photonic memory wires, is placed in between two resonant rings (**Fig. 2b**) to selects the opportune wavelength front and back, respectively (**Fig. 3a**). That is, once the PCM memory states are set (WRITE operation) in this photonic kernel (i.e. matrix ***B***), this architecture allows performing the weighting functionality entirely passive. Selective 'writing' is achieved by changing the phase of the corresponding number of PCM wires that we deposited on the waveguides, by local electrostatic heating, which promotes crystallization or amorphization, and consequently modifies the waveguide modal refractive index in a reversible process.

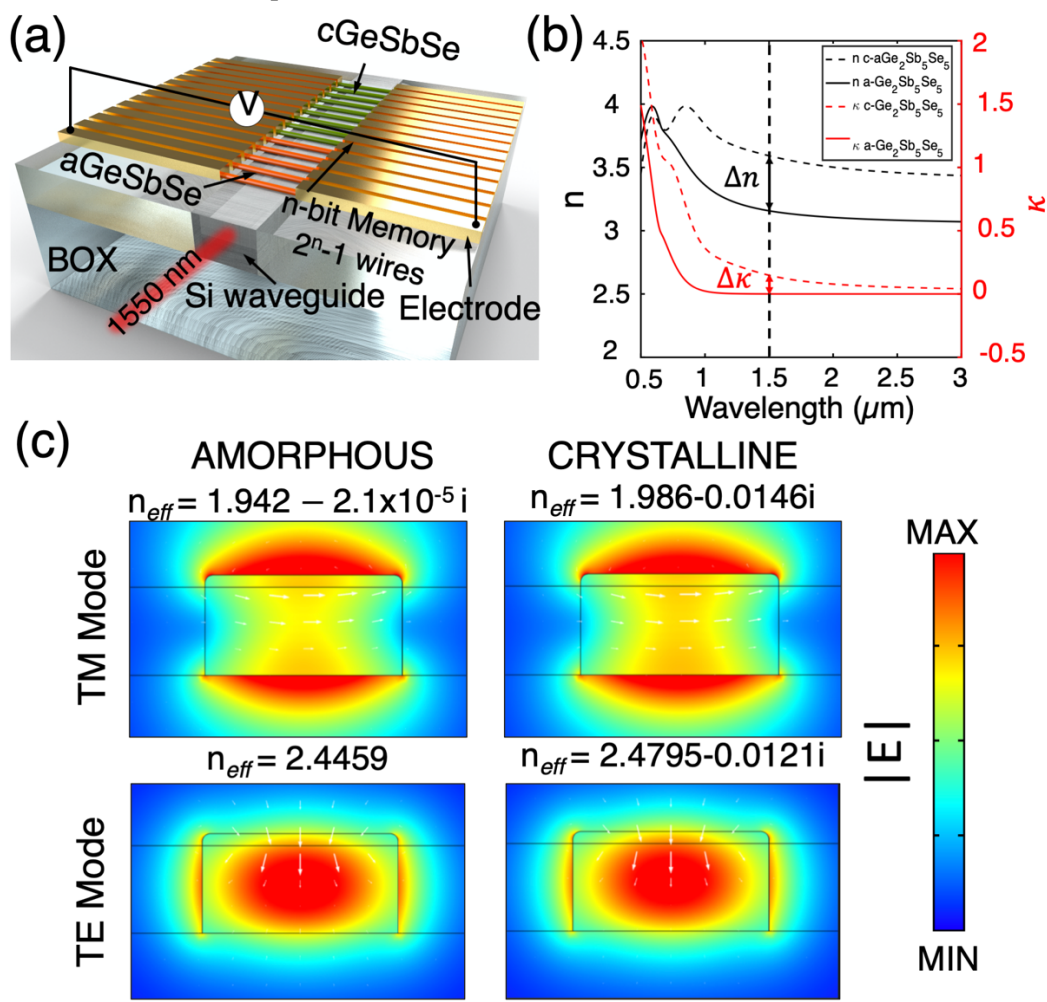

***Figure 3. Multistate on-chip photonic memory embedded in the PTC.** (**a**) Schematic representation of the multistate reprogrammable photonic memory. 30 nm thin and 250 nm wide $Ge_2Sb_2Se_5$ wires can be patterned using a combination of lithographic process, sputtering and liftoff. The resulting wire can be contacted with tungsten electrodes which represents the heating elements. A n-bit photonic memory of $2^n$ states require $2^n-1$ wires. (**b**) Experimentally obtained (ellipsometry) optical properties of phase change material ($Ge_2Sb_2Se_5$) film. Real (n, left y-axis) and imaginary (κ, right y-axis) parts of the refractive indices of the amorphous (solid line) and crystalline alloys. (dashed line). The $Ge_2Sb_2Se_5$ shows a sensitive variation of the absorption coefficient, while simultaneously showing small induced loss. (**c**) Simulated interaction between the $Ge_2Sb_2Se_5$ nanowire and the fundamental optical mode for both the amorphous (left panel) and crystalline (right panel) state. Fundamental transversal electric (TE) and Transverse magnetic mode profiles (normalized electric field) of the $Ge_2Sb_2Se_5$-Silicon hybrid waveguide at 1550 nm for amorphous and crystalline state show a strong index (imaginary-part) difference ~ 0.01, while incurring a relatively low insertion loss. White arrows represent the direction and intensity of the magnetic field ($H_x,H_y$).*

We decide to implement the photonic memory kernels based on $Ge_2Sb_2Se_5$, since this material presents broadband transparent region for telecommunication wavelengths in its amorphous state and can be used to implement high-performance nonvolatile multistate photonic memories.[32] $Ge_2Sb_2Se_5$ exhibits 3-orders of magnitude lower absorption coefficient with respect to regularly employed GST at 1550 nm, and features still a high optical (real part) index contrast Δ*n* of 0.5 across the near- to mid-IR bands and around 0.2 Δ*κ* in the C-band (**Fig. 3b**)[33]. Remarkably, the optical absorption in the amorphous state is vanishingly small and non-measurable when heterogeneously integrated in silicon photonics of ~100 micrometer long lengths. Moreover, the relatively lower variation of the absorption coefficient, indeed, makes it a promising material for multistate devices, avoiding the utilization of high laser power and extremely low noise equivalent power detectors. Assuming a continuous film, for the fundamental TM mode of the waveguide the phase transition produces a variation of the effective absorption coefficient Δκ~0.01 to which corresponds 0.21dB/μm (**Fig. 3c**).

When the network is trained, the extracted weights are set by changing through electrothermal switching individual states of photonic memories, instead of the previously used optical pulses[34]. Each state of the memory can reversibly be written, by selective transitioning between amorphous (*a*) and crystalline (*c*) phase using electrothermal switching induced by Joule heating, as previously demonstrated[35] and reported in **Fig. 4**.

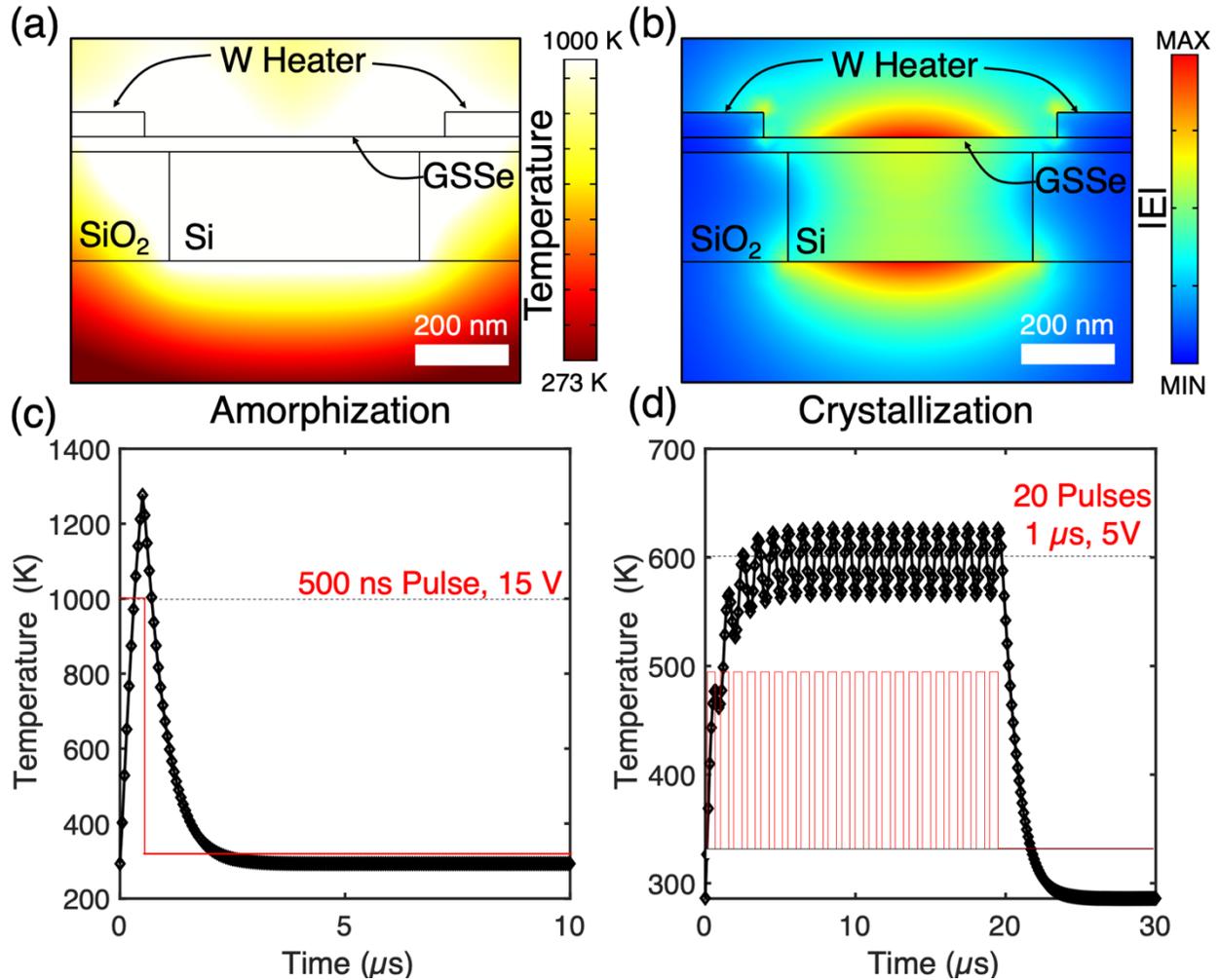

*Figure 4. Programming photonic memories using electrothermal switching. (a) Numerical study (COMSOL) of the electro-thermal switching at the equilibrium, heat map produced by Joule heating of a tungsten (W) and (b) normalized electric field mode profile of hybrid Si-GSSe waveguide. GSSe thickness = 30 nm. Time-trace of electro-thermal switching from; (c) the crystalline-to amorphous state and (d) vice versa using an electro-thermal mechanism. Average temperature in the GSSe layer vs. time (black dots) and corresponding electric signal applied to the thermal heaters for achieving it (red line). The amorphization temperature is equal (or larger) than the melting temperature of the film (>900K), while for crystallization it is below the amorphization temperature (~523K) and kept approximately constant for 20μs.*

In our scheme heat is applied to the material externally via joule heating of a tungsten metal layer in contact with the wire, as shown in **Fig. 4a**, whose mode profile and thermal profile is simulated in **Fig. 4a-b**. Different pulse train profiles according to the type of transition (*a-c* or *c-a*) are applied to the wire via the connected in series to the device.[35] (**Fig. 4 c-d**).

The material choice for the electrodes, their placement with respect to the waveguide and the propagating mode are opportunely engineered to minimize the insertion losses while providing efficient thermal energy. This was possible because the metal of choice (tungsten, W) has superior thermal properties without being affected by high optical losses such as in plasmonic noble metals. Moreover, the periodicity and intensity of the electric pulses applied to the tungsten electrodes, used for writing the memory, has to be adjusted for providing sufficient thermal energy to $Ge_2Sb_2Se_5$ wire according to in which phase has to be switched. The voltage, number of pulses and periodicity can be regulated to i) heat up the PCM wire up to 250°C and anneal for few tens of micro seconds to crystallize it  ii) melt it, increasing the temperature to over 600 °C, for the amorphization.[34,35] A resistive heater optimized for efficient switching and contemporary not generating insertion losses, can also be made in doped silicon or in silicide, currently used in p-n modulator[36], positioned next to the waveguide, ITO or graphene electrodes[37].

Light signals that couple with this phase-change memory probe the variation of the absorption coefficient over phase transition (READ operation). Considering a total of 16 states (4-bit), the 30 nm thin and 250 nm wide programable PCM-wires arranged in a grating fashion (duty cycle 50%), it is possible to obtain a 4-bit memory for each element of the kernel ($B_{ij}$), with a total length of just ~8 μm, excluding electrical circuitry (**Fig. 5a**). The insertion losses, defined as $10\log_{10}\left(\frac{P_0}{P_{input}}\right)$ where $P_0$ is the optical power transmitted when all the wires are in the amorphous state, is only ~1dB for a 4-bit multilevel memory (**Fig. 5a – i**). The optical power transmitted decreases when the GSSe wires are written (switching to crystalline) leading to discrete power levels for each quantized state (**Fig. 5a – ii-iii**). The insertion losses for multistate memory devices with different quantization resolution (1-bit through 4-bit) are shown in **Figure 5b** and are derived from numerical simulations reported in the supplementary materials **(Figure S2)**, which highlights the electric field distribution. The photonic memory implemented in this configuration provides a uniform quantization. For a 4-bit photonic memory, the quantization step is 0.2 dB/bit and a maximum extinction ratio of about 3.5 dB of modulation depth (**Fig. 5c**). Here the extinction ratio is computed as the ratio of the optical power transmitted in the 2 extreme configurations: all the wires in the crystalline state and all the wires in the amorphous state, $10\log_{10}\left(\frac{P_{1111}}{P_{0000}}\right)$.

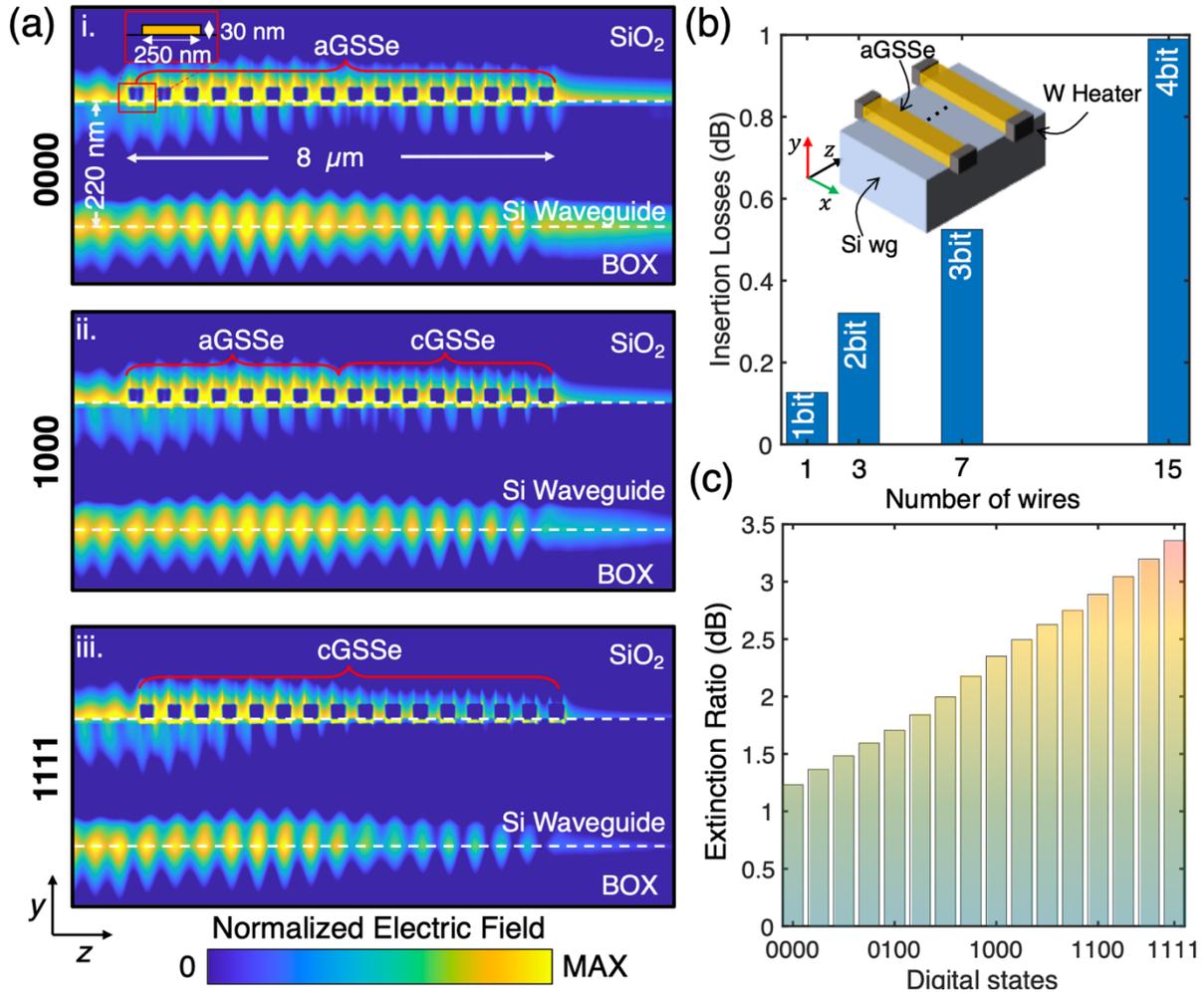

*Figure 5. 4-bit multistate on-chip reprogrammable photonic memory. (a) Normalized electric field in the yz plane of a propagating wave (TM mode) for different digital states (from up to bottom i. '0000', ii. '1000' and iii. '1111'). Here we considered the 0000 state to be less lossy and 1111 to be more lossy. Dashed lines represent top and bottom waveguide surfaces. Dimension of waveguides, GSSe wire are reported for state 0000. (b) Insertion loss (IL) considering array of different number of wires of amorphous GSSe, which represents insertion losses of photonic memory of different states; 1 PCM bar = 2 states (1-bit); 3 PCM bars = 4 states (2-bit); N bars = N+1 states = $Log_2(N+1)$ bits. Note, optical losses include also the tungsten electrodes to WRITE/RESET the photonic memories. (c) Extinction ratio (ER) for a 4-bit photonic memory as function of digital states, for an increased number of crystalline-wire the ER increases linearly and uniformly, giving an ER/IL = 3.5. Details of the numerical simulation and preliminary experimental simulations are given in the Supplementary Material Section S1 and S2, respectively.*

The functionality of the dot-product engine, which is the building block of the proposed PTC unit, along with its programmability is assessed using circuit level interconnect simulations (**Fig. 6a**). For evaluating the engine's performance, we simulate a dot product between two 4-elements vectors: input unitary row-vector $a_{1,i}$ and the column-vector $b_{i,1}$, with the latter being stored in the photonic memory. For illustrative purposes, we considered the last element of $b_{i,1}$ (i = 4) to be updated using a pseudo-random switching 1) varying the content of the photonic memory from the lowest state to the highest (**Fig. 6b**) by parallelly switching all the PCM wires and 2) varying just the Least Significant Bit (LSB) (**Fig. 6c**). Bearing in mind the relatively slow and asymmetric switching speed, the moderately low noise equivalent power (NEP) of the photodetector and overall low spectral noise, it is possible to well discriminate (eye-diagram completely open) results of the dot product between numbers which differ by the smallest possible increment, of namely 1-bit. It is worth mentioning that the discrimination between states can be aided by varying the dynamic range of the photonic memory, trading off insertion losses, e.g. using wider PCM wires. (Further details on the interconnect simulations are discussed in the Supplementary Materials, Section S3)

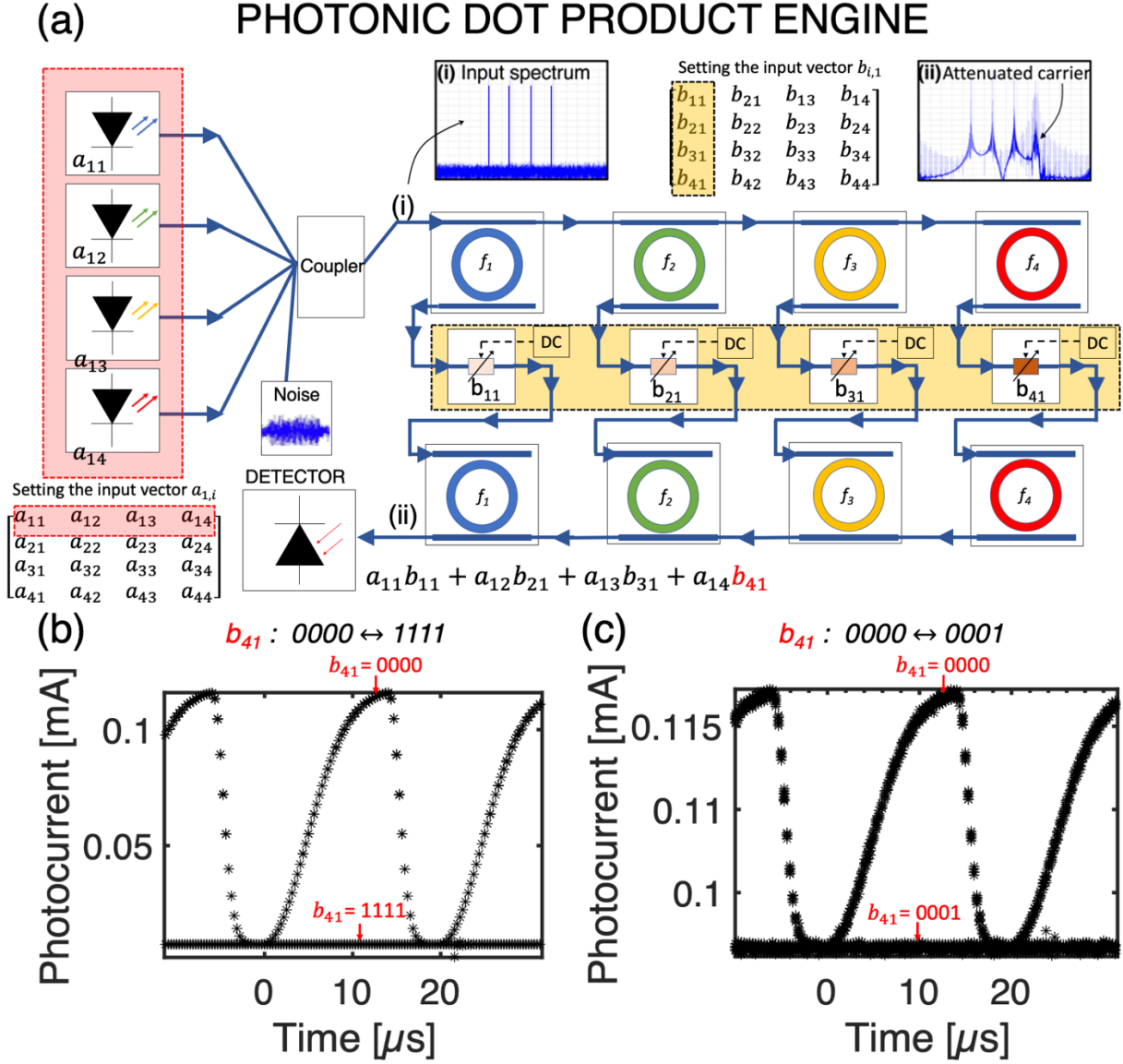

*Figure 6. Dot product engine interconnect simulation. (a) Layout of the schematic used for the interconnect simulation for testing the dot-product engine. Inset (i) represents the input spectrum with additional noise (Spectral Density = $10^{-18}$ W/Hz). Inset (ii) Spectrum after intensity filtering assuming $b_{41}$= "1111" and all the other element of the matrix $b_{ij}$=0000 (low attenuation). The photonic memories are modelled as modulators with tabular quantized states, the thermal heater are used for setting the state of the photonic memories and the corresponding extinction ration (ER) and a photodetector, characterized by low dark current (50 nA), affected by both thermal and shot noise, detect the incoherent summation. (b-c) Eye diagrams for a dot product in which only one element of the vector changes all bits (b, from 1111 → 0000) or the least Significant bit only (c, 0000 to 0001).*

## II.4 Performances

The PTC implemented according the proposed scheme can perform matrix multiplication with 4-bit precision, completely passively once the weights are stored in the photonic network, which is a one-time operation. The PTC does not rely on any logic architecture nor requires transduction from off-chip memory when performing inference, therefore it could be considered a full-fledged analog processor, as other recently developed[10,38]. In fact, during inference tasks, when our architecture performs tensor operations (*More details are provided in the Supplementary Materials Section S4*) with a time complexity of *O*(1) and the static power consumption is approaching zero, since the system behaves as a passive filter and simply relies on light matter interactions with pre-stored states in the photonic memory (kernel have been already saved in the photonic memory in a former instance, and the inputs are readily accessible from the optical domain, assuming being situated at the edge-of the network) instead of logic operations which requires optical switching[39,40].

An initial performance analysis is as follows: considering a photonic foundry Ge-photodetectors, micro ring resonator (radius = 10 μm) and AIM-photonics disc-modulators, the latency of an individual PTC sub-unit (e.g. unit $D_{2,1}$) requires Σ{Electro-Optic (E2O) + Time-of-Flight (ToF) + Detection (Rx)} = ~65 ps for processing a 4x4 matrix multiplication resulting in computing 64 MACs at 4 bit precision. This delivers a total 0.5-2 Peta operations per second (POPS/s) throughput (assuming layout of **Fig. 2**, and 1 MAC = 2 Operations – OPS) for ~250 4x4 PTC units, when limiting the maximum die-area to 800 mm$^2$ (4-bit Digital to Analog Converters – DAC – area = 0.05 mm$^2$) limited mainly by the Electro-to-Optical conversion (i.e. DACs). For an optical data input (e.g. camera), the peak throughput increases to 16 POPS/s for only a few watts of power. If pipelining could be used, the 65 ps drops to ~20 ps latency, thus improving throughputs by another 3x, and hence one could consider sharing DAC usage amongst cores. Another key aspect to consider is the overall power consumption, which considering the completely passive operation of the dot-product and accumulation (and zero bias in detection with high responsivity photoreceivers), accounts only for the laser power which is below 5mW and bias of the photodetector in case for the optical data input (middle column, **Table I**), but also the E2O DACs in the electronic-data input case (left column, **Table I**). Although, it is important to mention that when performing inference, thanks to robustness of the neural network achievable through opportune training, low-bit quantization of the weights is also possible, obtaining indeed efficient and accurate inference for low resolution quantized weights.[41,42] If the system would be used to perform relatively simple inference tasks at the edge of a network it may not require a high bit resolution. It is worth mentioning that for GPUs and digital architectures, a great portion of the power consumption is related to other tasks performed and off-chip memory. For this reason, we believe that PTC, and in general neuromorphic photonics, including emerging opto-electronic technology[43-48], is advantageous with respect to digital electronics only for specific applications that can exploit (1) inherent operations utilizing efficient light matter interactions, such as the one proposed in this research, avoiding cumbersome domain conversion (lowering overall power consumption) (2) that would require ns-fast inference. In this view, the proposed photonic engine can be used for accelerating inference tasks for specific applications in which obtaining a prediction (even at reduced accuracy) in near real-time (~ns delay) is essential and has the priority over obtaining a prediction with high accuracy with longer latency. We particularly envision its use for tasks in which the input data are already RF or optical signals avoiding inefficient analog-to-digital conversion." Additionally, we also compared the potential performances of the PTC with the performance achieved by the state-of-the-art GPU (Nvidia A100, May 2020) when performing tensor operations at 4-bit. The A100 requires a total max power of 400 W, which is 1 order of magnitude higher than the T4, featuring a higher number of tensor cores. Remarkably, the total throughput is still 1 order of magnitude lower compared the potential throughput provided by the photonic tensor core when performing inference (when no writing operation is performed).

TABLE I. Tensor ore performance comparison. Electronic data-fed (left column) Photonic Tensor Core (PTC) offers 2-8x throughput improvement over NVIDIA's T4 and A100 and for optical data (e.g. camera) improvements are ~60x (chip area limited to a single die ~800mm$^2$). *10:1 DAC (Digital to Analog converter) reuse. ** Optical Data input (no DACs). ***Inference only.

|  | Electronic Data PTC | Optical Data PTC** | NVIDIA T4*** | A100 |
| --- | --- | --- | --- | --- |
| # of Tensor Cores | 250 | 250 | 320 | 512 |
| Clock Speed | 50 GHz | N.A. | <1.5 GHz | <1.5 GHz |
| Bit resolution | 4-bit | 4-bit | 4-bit | 4-bit |
| Throughput (POPS/s) | 0.5 (~2)* | ~16 | 0.26 | 1.26 |
| Power | 81W | < 2W | 70W | 400W (Max) |
| Operation Efficiency (TOPS/J) | 25 | ~10$^3$ | 3 | 4 |

## III. CONCLUSION

In summary, we propose a tensor core unit implemented in photonics that relies on photonic multiplexed (WDM) signals, weighted, after filtering, using engineered multi-state photonic memories based on $Ge_2Sb_2Se_5$ wires patterned on the waveguide. The photonic memories are reprogrammed by selectively changing phase (amorphous/crystalline) of the wires, using electrothermal switching through Joule heating induced by tungsten electrodes. The photonic memory programming can be realized in parallel (few microseconds), if needed, or alternatively, this photonic tensor core can operate as a passive system with a pre-SET kernel matrix; that is there will be no dynamic nor static power dissipation. The runtime complexity is thence $O(1)$. An additional key technology feature of this design is that no additional losses are introduced by the photonic memories, avoiding repeaters, optical amplifiers cumbersome EO/OE conversions. The architecture shows execution time limited only by the time of flight of the photon in the chip, which is function of the ring size/selectivity (number of wavelengths), and the latency of the photodetector $O(<10^{-1}$ns) once the kernel matrix is set and optical input data is being processed. The concurrent development of new PCM materials and the advancement of the integration of photonic memories can enable the realization of engines based on the proposed scheme able to inherently perform full precision floating point matrix multiplication and accumulation, and consequently opening a pathway towards the realization of all-optical photonic tensor units which can significantly speed up intelligent tasks at the edge of the network without requiring EO conversions and access to external memories.

## ACKNOWLEDGMENTS

We acknowledge the support from the Presidential Early Career Award for Scientist and Engineers (PECASE) nominated by the Department of Defense namely by the Air Force Office of Scientific Research. The authors would like to thank Prof. A. Kildishev and Prof. Juejun Hu for the insightful discussion.

### Availability of the Data

The data that support the findings of this study are available from the corresponding author upon reasonable request.